\documentclass[aps,prd,twocolumn,superscriptaddress,showpacs]{revtex4}
\usepackage{epsfig,epsf}
\usepackage{amsmath}
\usepackage{amsthm}
\usepackage{amsfonts}
\usepackage{amssymb}
\usepackage{dsfont}
\usepackage{multirow}
\usepackage{appendix}
\usepackage{slashed}
\usepackage[active]{srcltx}
\usepackage{psfrag}

\begin{document}

\title{Application of the QCD light cone sum rule to tetraquarks: the strong
vertices $X_bX_b\rho$ and $X_cX_c\rho$}
\date{\today}
\author{S.~S.~Agaev}
\affiliation{Department of Physics, Kocaeli University, 41380 Izmit, Turkey}
\affiliation{Institute for Physical Problems, Baku State University, Az--1148 Baku,
Azerbaijan}
\author{K.~Azizi}
\affiliation{Department of Physics, Do\v{g}u\c{s} University, Acibadem-Kadik\"{o}y, 34722
Istanbul, Turkey}
\author{H.~Sundu}
\affiliation{Department of Physics, Kocaeli University, 41380 Izmit, Turkey}

\begin{abstract}
The full version of QCD light-cone sum rule method is applied to tetraquarks
containing a single heavy $b$ or $c$ quark. To this end, investigations of
the strong vertices $X_{b}X_{b}\rho$ and $X_{c}X_{c}\rho$ are performed, where $X_b=[su][\bar b\bar d]$
and $X_c=[su][\bar c\bar d]$ are the exotic states built of  four quarks
of different flavors. The strong coupling constants
$G_{X_{b}X_{b}\rho}$ and $G_{X_{c}X_{c}\rho}$ corresponding to these vertices are found using
the $\rho$-meson leading and higher-twist distribution amplitudes. In
the calculations $X_{b}$ and $X_{c}$ are treated as scalar bound states of a diquark and
antidiquark.
\end{abstract}

\maketitle

\section{Introduction}

During last decade, due to experimental data of the Belle, BaBar, LHCb, D0
and BES collaborations, which provided valuable information on the so-called
exotic hadron states, this branch of high energy physics demonstrates a
rapid growth. The exotic hadrons, i.e. ones that cannot be embraced by the
spectroscopy of the known hadrons as $q\bar q$ or $qqq$ bound states, may
serve as a laboratory for testing the Quantum Chromodynamics (QCD)- the
existing theory of strong interactions, as well as various phenomenological
models built of on its basis. An existence of the exotic hadrons does not
contradict the fundamental principles of this theory. Though relevant
problems attracted an interest of physicists from first years of the parton
model and, later QCD, only recently these ideas found their experimental
confirmation.

The discovery of the charmonium-like resonance $X(3872)$ by the Belle
Collaboration \cite{Choi:2003ue} was the first brick laid on footing of the
house, which now exists as XYZ family of the exotic states. The observation
made by the Belle was later reexamined and confirmed by other collaborations
\cite{Abazov:2004kp,Acosta:2003zx,Aubert:2004ns}. Produced in the $B$ meson
decays or in the $pp$ collisions, observed in the $e^{+}e^{-} $ annihilation
or in the two-photon fusion, exotic states remain on the focus of the main
experimental collaborations, which collected wide data base on the processes
of interest.

A considerable progress was made in the theoretical understanding of the
features of the exotic states, as well. If experiments are devoted to
measuring of the masses, and decay widths, to identifying the spins and
parities of the exotic states, theoretical works are concentrated on studies
of their internal quark-gluon structure, on new models and methods suggested
for their exploration (for details of theoretical and experimental studies see, the reviews \cite{Jaffe:2004ph,Swanson:2006st,Klempt:2007cp,Godfrey:2008nc,Voloshin:2007dx,Nielsen:2009uh,
Faccini:2012pj,Esposito:2014rxa,Chen:2016}, and references therein).

The charmonium-like resonances of the XYZ family contain, as it is evident
from their names, a $c \bar c$ component. Therefore, efforts were done to
explain the new resonances as excitations of the ordinary $c\overline{c}$
charmonium. Indeed, some of new particles allow such interpretation, and are
really excited $c\overline{c}$ states. But the essential part of the
relevant experimental data cannot be included into the exited charmonium
scheme, and hence for their exploration unconventional quark-gluon
configurations are needed. For this purpose, various models with different
quark-gluon structures were supposed. The tetraquark model of the exotic
states, i.e. the model that considers exotics as the four-quark particles,
is among mostly employed ones. It is worth to note that this approach led to
significant achievements in describing of the processes with the exotic
states, in predicting their masses, decay widths and quantum numbers. There
are some alternatives to compose from the four quarks an exotic state within
the tetraquark model. In fact, the four constituent quarks may group into a
diquark and an antidiquark to form the exotic state with required quantum
numbers. This model is known as the diquark-antidiquark model. In the meson
molecule picture the quarks are collected into two conventional mesons, and the
exotic particle appears as loosely-bound molecule state. There are other
opportunities to organize the exotic states from the four quarks, as well as
alternative models, for an example, the hybrid models detailed presentation
of which is beyond the scope of the present work.

In the tetraquark model the maximal number of the quark flavors in the XYZ states
does not exceed three. But there are not any fundamental laws
in QCD forbidding the existence of the exotic states built of four quarks of
distinct flavors. Namely such exotic states recently became the objects of
comprehensive theoretical investigations. But before going into details of
these studies, we have to make some comments on the experimental situation
formed around one of such particles. Strictly speaking, all present
theoretical activity was inspired by the D0 collaboration's report, where
an evidence for existence  of the exotic state $X(5568)$ was announced \cite{D0:2016mwd}.
Based on analysis of $p\bar{p}$ collision data at $\sqrt{s}=1.96\ \mathrm{%
TeV}$ collected at the Fermilab Tevatron collider, the collaboration
reported on evidence of a narrow resonance $X(5568)$ in the consecutive
decays $X(5568) \to B_{s}^{0} \pi^{\pm}$, $B_{s}^{0} \to J/\psi \phi$, $%
J/\psi \to \mu^{+} \mu^{-}$, $\phi \to K^{+}K^{-}$. From the  decay
channel $X(5568) \to B_{s}^{0} \pi^{\pm} $ it is easy to conclude that the
state $X(5568)$ consists of valence $b,\, s, \, u$ and $d$ quarks. The mass
of this state is equal to $m_{X}=5567.8 \pm 2.9 \mathrm{(stat)}%
^{+0.9}_{-1.9} \mathrm{(syst)}\, \mathrm{MeV}$, and decay width is estimated
as $\Gamma=21.9 \pm 6.4 \mathrm{(stat)}^{+5.0}_{-2.5} \mathrm{(syst)}\,
\mathrm{MeV}$. The D0 assigned to this particle the quantum numbers $%
J^{PC}=0^{++}$, but did not exclude also a possible version $1^{++}$. Few
days later the LHCb Collaboration presented preliminary results of their
analysis of $pp$ collision data at energies $7\, \mathrm{TeV}$ and $8\,
\mathrm{TeV}$ collected at CERN \cite{LHCb:2016}. The LHCb Collaboration
could not confirm the existence of the resonance structure in the $%
B_s^{0}\pi^{\pm} $ invariant mass distribution at the energies less than $%
5700\ \, \mathrm{MeV}$. In other words, situation with the exotic state $%
X(5568)$, supposedly built of four different quark flavors is controversial
and necessitates further experimental studies. The exotic state dubbed $%
X(5568)$ deserves to be searched for by other collaborations, and maybe, in
other hadronic processes.

Namely these unclear circumstances surrounding  the  $X(5568)$ resonance
make relevant theoretical studies even more important than just
after the information on its existence. First suggestions concerning
the diquark-antidiquark or meson molecule model for organization of the new state
were made in Ref.\ \cite{D0:2016mwd}. Calculations performed until now covered only some topics
of the $X(5568)$ physics. They include mainly computation of the mass, decay
constant of $X(5568)$; a few works were devoted to
calculation of the width of the $X(5568) \to B_{s}^{0} \pi^{\pm}$ decay, as
well. It should be emphasized that the diquark-antidiquark model with $%
J^{PC}=0^{++}$ prevails among approaches used to explain parameters of the $%
X(5568)$ state.

Thus, in Ref.\ \cite{Agaev:2016mjb} we accepted for this state the
diquark-antidiquark structure $X_b=[su][\bar{b}\bar{d}]$ with the quantum
numbers $0^{++}$, and calculated its mass $m_{X_b}$ and decay constant
(i.e. the meson-current coupling) $f_{X_b}$. Our prediction for $m_{X_b}$ agrees with
the mass of the $X(5568)$ resonance found by the D0 collaboration. In the framework of the
diquark-antidiquark model some parameters of $X(5568)$ were also analyzed in
Refs.\ \cite{Wang:2016tsi,Chen:2016mqt,Zanetti:2016wjn,Wang:2016mee}, where
an alternative choice for the diquark-antidiquark type interpolating current
was realized. The values for $m_{X}$ obtained in these works agree with each
other, and are consistent with the experimental data of D0 Collaboration.

Employing the same $X_b$ structure and interpolating current as in our
previous work, in Ref.\ \cite{Agaev:2016ijz} we computed the width of the $%
X_b \to B_{s}^{0}\pi^{+}$ decay channel. We applied QCD sum rule on the
light-cone supplemented by the soft-meson approximation (see, Ref.\ \cite%
{Agaev:2016dev}): our result for $\Gamma(X_{b}^{+}\to B_{s}^{0}\pi^{+})$
describes correctly the experimental data. The width of the decay channels $%
X^{\pm}(5568) \to B_{s}\pi^{\pm}$ was also calculated in Refs.\ \cite%
{Dias:2016dme,Wang:2016wkj} using the three-point QCD sum rule approach. In
these works authors found a very nice agreement between the theoretical
predictions for $\Gamma(X^{\pm} \to B_{s}^{0}\pi^{\pm})$ and the data.

The $X(5568)$ can also be considered as a meson molecule; namely this
picture was realized in Refs.\ \cite{Xiao:2016mho,Agaev:2016urs}, where $%
X(5568)$ was treating as the $B\overline{K}$ bound state. It is worth to
note that, in accordance with Ref.\ \cite{Agaev:2016urs}, the mass of such
molecule-like state was found equal to $m_{X_{b}}=5757 \pm 145\, \mathrm{MeV}
$.

A charmed partner of the $X_b$ state, i.e. the $X_c$ structure built of the
valence $c,\,s,\, u$ and $d$ quarks and possessing the quantum numbers $%
0^{++}$ was analyzed in Ref.\ \cite{Agaev:2016lkl}. Here, we computed the
mass, decay constant and width of the decays $X_c \to D_{s}^{-}\pi^{+}$ and
$X_c \to D^{0}K^{0}$ considering $X_c=[su][\bar{c}\bar{d}]$ as the
diquark-antidiquark state and employing two forms for the interpolating
currents. The questions of quark-antiquark organization of $X_b$ and its
partners were also addressed in Ref.\ \cite{Liu:2016ogz}.

The contradictory information by the D0 and LHCb collaborations concerning
existence of the $X_b$ state resulted in an appearance of interesting
theoretical works devoted to analysis of the $X_b$ physics, where its structure and
spectroscopic parameters, production mechanisms were investigated. For details and further
explanations we refer to original papers \cite%
{He:2016yhd,Jin:2016cpv,Stancu:2016sfd,Burns:2016gvy,Tang:2016pcf,Guo:2016nhb,Lu:2016zhe, Esposito:2016itg,Albaladejo:2016eps,Ali:2016gdg}.

In the present work we explore the strong vertices $X_{b}X_{b}\rho
$ and $X_{c}X_{c}\rho $, and calculate the couplings $G_{X_{b}X_{b}\rho}$ and
$G_{X_{c}X_{c}\rho}$ by employing the QCD light-cone sum rule (LCSR) approach, which is one of
the powerful nonperturbative methods in  hadron physics enabling us to evaluate
parameters of the particles and processes \cite{Balitsky:1989ry}. Within this approach
one expresses the relevant correlation functions as convolution integrals of the
perturbatively calculable coefficients and non-local matrix elements, which
are the distribution amplitudes (DAs) of the particles under consideration.
It is worth noting, that expansion in terms of non-local matrix elements
cures shortcomings of the local expansion used in the conventional
QCD sum rules.

Strictly speaking, the light cone expansion was already applied for
investigation of the exotic states. Indeed, in order to study strong
vertices involving the exotic states, and calculate corresponding couplings
and decay widths in Refs.\ \cite{Agaev:2016ijz,Agaev:2016dev,Agaev:2016urs}
we applied a technique of the light cone calculations and obtained very good
results, which agree with available experimental data and predictions of other theoretical works.
But because of the differences in the quark contents of the conventional and exotic mesons,
in those works we had to supply the light cone expansion by the soft-meson approximation; the latter
reduces the light cone expansion to the expansion in terms of local matrix
elements weakening effects and advantages of the LCSR.

In the present work we employ the full version of the LCSR method in
computation of the strong vertex composed of the exotic particles. This
method previously was applied to analyze numerous vertices of conventional
mesons and baryons, and calculate corresponding couplings, form factors.
Here we are able to cite only some of the  works devoted to this
interesting topic of hadron physics \cite{Belyaev:1994zk,Aliev:1996xb,Khodjamirian:1999hb,Aliev:2010yx,Aliev:2011ufa,Agaev:2015faa}
noting among them Ref.\ \cite{Agaev:2015faa}, where, for the first time, effects of
the $\eta$ and $\eta^{\prime}$ mesons' gluon components on the
strong vertices $D^{*}_sD_{s}\eta^{(\prime)}$ and $B^{*}_sB_{s}\eta^{(%
\prime)}$  were taken into account. To our best knowledge, the present work is the first
attempt to investigate the strong vertex of tetraquarks by employing
the full version of QCD LCSR method. Therefore, it is instructive to reveal
possible technical problems hidden behind such kind of calculations, and
elaborate schemes and methods to evade them.

This work is structured in the following manner. In Sect.\ \ref{sec:SumRule}
we derive the light-cone sum rule for the strong coupling $%
G_{X_{b}X_{b}\rho} $ using the expansion of the correlation function
in terms of the $\rho$ meson's two- and three particle distribution amplitudes of various twists. In
Sect.\ \ref{sec:Num} we perform numerical analysis of the obtained sum rules for the couplings
$G_{X_{b}X_{b}\rho}$ and $G_{X_{c}X_{c}\rho}$. Appendixes \ref{sec:App1} and \ref{sec:App2}
contain some technical details of calculations and formulas useful in the continuum subtraction,
respectively.

\section{Sum rule for the coupling $G_{X_{b}X_{b}\protect\rho}$}
\label{sec:SumRule}
In this section we derive the sum rule for the strong
coupling $G_{X_{b}X_{b}\rho }$; the same expressions, after trivial
replacements of the meson and quark masses, can be applied for computation
of the coupling $G_{X_{c}X_{c}\rho }$, as well.

To calculate the coupling $G_{X_{b}X_{b}\rho }$ corresponding to the vertex $%
X_{b}X_{b}\rho $ in the framework of the QCD light-cone sum rules method, we
consider the corresponding correlation function, which in the case under
consideration is given by the expression
\begin{equation}
\Pi (p,q)=i\int d^{4}xe^{ipx}\langle \rho (q)|T\left\{
J^{X_{b}}(x)J^{X_{b}\dagger }(0)|0\right\} \rangle  \label{eq:CF1}
\end{equation}%
where $J^{X_{b}}(x)$ is the current with required quantum numbers within the
diquark-antidiquark model of the $X_{b}$ state defined in the form
\begin{equation}
J^{X_{b}}(x)=\varepsilon ^{abc}\varepsilon ^{ade}\left[ s^{b}(x)C\gamma
_{5}u^{c}(x)\right] \left[ \overline{b}^{d}(x)\gamma _{5}C\overline{d}^{e}(x)%
\right] .  \label{eq:CR1}
\end{equation}%
First, let us calculate this function in terms of the physical degrees of
freedom. We get
\begin{eqnarray}
\Pi ^{\mathrm{Phys}}(p,q) &=&\frac{\langle 0|J^{X_{b}}|X_{b}(p)\rangle }{%
p^{2}-m_{X_{b}}^{2}}\langle \rho (q)X_{b}(p)|X_{b}(p+q)\rangle  \notag \\
&&\times \frac{\langle X_{b}(p+q)|J^{X_{b}\dagger }|0\rangle }{%
(p+q)^{2}-m_{X_{b}}^{2}}.
\end{eqnarray}%
Here the matrix element $\langle \rho (q)X_{b}(p)|X_{b}(p+q)\rangle $
determines the coupling of interest and is given as
\begin{equation}
\langle \rho (q)X_{b}(p)|X_{b}(p+q)\rangle =G_{X_{b}X_{b}\rho }p\cdot
\varepsilon ,
\end{equation}%
where $p$ is the momentum of the $X_{b}$ state, and $\varepsilon ^{\mu }$ --
polarization vector of the $\rho $-meson. We define also by the standard
manner the matrix element
\begin{equation}
\langle 0|J^{X_{b}}|X_{b}(p)\rangle =m_{X_{b}}f_{X_{b}}.
\end{equation}%
Then we easily find%
\begin{eqnarray}
&&\Pi ^{\mathrm{Phys}}(p,q)=\frac{m_{X_{b}}^{2}f_{X_{b}}^{2}G_{X_{b}X_{b}\rho }%
}{(p^{2}-m_{X_{b}}^{2})\left[ (p+q)^{2}-m_{X_{b}}^{2}\right] }p\cdot
\varepsilon \notag \\
&&+\ldots
\end{eqnarray}%
where the first term is the ground state contribution and dots stand for the contributions
arising from the higher resonances and continuum states. As is seen, the correlation
function contains only the structure $p\cdot\varepsilon $. The relevant invariant amplitude
is given by the expression
\begin{eqnarray}
&&\Pi ^{\mathrm{Phys}}(p^{2},(p+q)^{2})=\frac{%
m_{X_{b}}^{2}f_{X_{b}}^{2}G_{X_{b}X_{b}\rho }}{(p^{2}-m_{X_{b}}^{2})\left[
(p+q)^{2}-m_{X_{b}}^{2}\right] } \notag \\
&&+\int \int \frac{ds_{1}ds_{2}\rho ^{\mathrm{phys}}(s_{1},s_{2})}{%
(s_{1}-p^{2})[s_{2}-(p+q)^{2}]} +\ldots
\end{eqnarray}%
Here the dots indicate the single dispersion integrals that
should be included to make the expression finite: they vanish after double Borel
transformations.

The Borel transformations on variables $p^{2}$ and $p^{\prime 2}=(p+q)^{2}$
applied to the invariant function yields
\begin{eqnarray}
&&\mathcal{B}_{p^{2}}(M_{1}^{2})\mathcal{B}_{p^{\prime 2}}(M_{2}^{2})\Pi ^{%
\mathrm{Phys}}(p^{2},p^{\prime 2})\equiv \Pi ^{\mathrm{Phys}}(M^{2}) \notag \\
&&=m_{X_{b}}^{2}f_{X_{b}}^{2}G_{X_bX _b\rho
}e^{-m_{X_{b}}^{2}/M^{2}} \notag \\
&&+\int ds_{1}ds_{2}e^{-{(s_{1}+s_2)}/2M^2}\rho ^{\mathrm{phys}}(s_{1},\,s_{2}),
\label{eq:BorT}
\end{eqnarray}%
where the Borel parameters $M_{1}^{2}$ and $M_{2}^{2}$ for the problem under consideration
are chosen as $M_{1}^{2}=M_{2}^{2}=2M^{2}$, and $M^{2}=M_{1}^{2}M_{2}^{2}/(M_{1}^{2}+M_{2}^{2})$.

To proceed we need to determine the correlation function using quark
propagators and distribution amplitudes of the $\rho $ meson, i.e. to find $%
\Pi ^{\mathrm{QCD}}(p,q)$. We note that it is the sum of two terms
\begin{equation*}
\Pi ^{\mathrm{QCD}}(p,q)=\Pi _{1}(p,q)+\Pi _{2}(p,q).
\end{equation*}%
The first function corresponds to a physical situation, when the strong
vertex is formed due to interaction of the $X_{b}$ states with the $%
\overline{d}d$ component of the $\rho ^{0}$ meson, and is determined by the
formula
\begin{eqnarray}
&&\Pi _{1}(p,q)=i\int d^{4}xe^{ipx}\varepsilon ^{abc}\varepsilon
^{ade}\varepsilon ^{a^{\prime }b^{\prime }c^{\prime }}\varepsilon
^{a^{\prime }d^{\prime }e^{\prime }}  \notag \\
&&\times \mathrm{Tr}\left[ \gamma _{5}\widetilde{S}_{s}^{b^{\prime
}b}(x)\gamma _{5}S_{u}^{cc^{\prime }}(x)\right] \left[ \gamma _{5}\widetilde{%
S}_{b}^{d^{\prime }d}(-x)\gamma _{5}\right] _{\alpha \beta }  \notag \\
&&\times \langle \rho (q)|\overline{d}_{\alpha }^{e}(x)d_{\beta }^{e^{\prime
}}(0)|0\rangle .  \label{eq:CF2}
\end{eqnarray}%
The second component $\Pi _{2}(p,q)$ appears via the interaction of the $%
X_{b}$ states and $\rho ^{0}$ meson's $\overline{u}u$ content:%
\begin{eqnarray}
&&\Pi _{2}(p,q)=-i\int d^{4}xe^{ipx}\varepsilon ^{abc}\varepsilon
^{ade}\varepsilon ^{a^{\prime }b^{\prime }c^{\prime }}\varepsilon
^{a^{\prime }d^{\prime }e^{\prime }}  \notag \\
&&\times \mathrm{Tr}\left[ \gamma _{5}\widetilde{S}_{d}^{e^{\prime
}e}(-x)\gamma _{5}S_{b}^{d^{\prime }d}(-x)\right] \left[ \gamma _{5}%
\widetilde{S}_{s}^{bb^{\prime }}(x)\gamma _{5}\right] _{\alpha \beta }
\notag \\
&&\times \langle \rho (q)|\overline{u}_{\alpha }^{c^{\prime }}(0)u_{\beta
}^{c}(x)|0\rangle .  \label{eq:CF3}
\end{eqnarray}%
In the equations above we introduce the notation%
\begin{equation*}
\widetilde{S}_{q,s,Q}(x)=CS_{q,s,Q}^{T}(x)C,
\end{equation*}%
where $S_{q(s,Q)}(x)$ are the quark propagators, and $C$ is the charge
conjugation matrix. In the $x$-space for propagators of the $u$,$d$ and $s$
quarks we accept the expressions
\begin{eqnarray}
&&S_{q}^{ab}(x)=\frac{i{\slashed x}}{2\pi ^{2}x^{4}}\delta _{ab}-\frac{m_{q}%
}{2\pi ^{2}x^{2}}\delta _{ab}  \notag \\
&&-ig_{s}\int_{0}^{1}dv\left\{ \frac{{\slashed x}}{16\pi ^{2}x^{2}}%
G_{ab}^{\mu \nu }(vx)\sigma ^{\mu \nu }-\frac{ivx^{\mu }}{4\pi ^{2}x^{2}}%
G_{ab}^{\mu \nu }(vx)\gamma ^{\nu }\right.  \notag \\
&&\left. -\frac{im_{s}}{32\pi ^{2}}G_{ab}^{\mu \nu }(vx)\sigma _{\mu \nu }%
\left[ \ln \left( -\frac{x^{2}\Lambda ^{2}}{4}\right) +2\gamma _{E}\right]
\right\} .  \label{eq:qprop}
\end{eqnarray}%
In Eq.\ (\ref{eq:qprop}) the first two terms are the perturbative components
of the propagator: terms $\sim G^{\mu \nu }$ appear due to its expansion on
the light-cone and describe interaction with the gluon field. In
calculations we neglect terms $\sim m_{q}$, and at the same time, take into
account ones $\sim m_{s}$. For the heavy quark propagator on the light-cone
we employ its expression in terms of the second kind Bessel functions $%
K_{\nu }(z)$
\begin{eqnarray}
&&S_{Q}^{ab}(x)=S_{Q}^{\mathrm{(0)}ab}(x)-\frac{g_{s}m_{Q}}{16\pi ^{2}}%
\int_{0}^{1}dv G_{ab}^{\mu \nu }(vx)\Bigg[ (\sigma _{\mu \nu }{\slashed x}%
 \notag \\
&&\left. +{\slashed x}\sigma _{\mu \nu })\frac{K_{1}\left( m_{Q}\sqrt{-x^{2}}%
\right) }{\sqrt{-x^{2}}}+2\sigma ^{\mu \nu }K_{0}\left( m_{Q}\sqrt{-x^{2}}%
\right) \right] ,  \notag \\
&&{}  \label{eq:Qprop}
\end{eqnarray}%
where the perturbative propagator of the heavy quark  is given by
\begin{eqnarray}
S_{Q}^{\mathrm{(0)}ab}(x) &=&\frac{m_{Q}^{2}}{4\pi ^{2}}\frac{K_{1}\left(
m_{Q}\sqrt{-x^{2}}\right) }{\sqrt{-x^{2}}}\delta _{ab}  \notag \\
&&+i\frac{m_{Q}^{2}}{4\pi ^{2}}\frac{{\slashed x}K_{2}\left( m_{Q}\sqrt{%
-x^{2}}\right) }{\left( \sqrt{-x^{2}}\right) ^{2}}\delta _{ab}.
\end{eqnarray}%
In Eqs.\ (\ref{eq:qprop}) and (\ref{eq:Qprop}) the shorthand notation
\begin{equation*}
G_{ab}^{\mu \nu }\equiv G_{A}^{\mu \nu }t_{ab}^{A},\,\,\,\,A=1,\,2\,\ldots 8,
\end{equation*}%
is adopted with $a,\,b$ being the color indices. Here $t^{A}=\lambda ^{A}/2$,
where $\lambda ^{A}$ are the Gell-Mann matrices.

\begin{figure}[h]
\includegraphics[width=5.5cm]{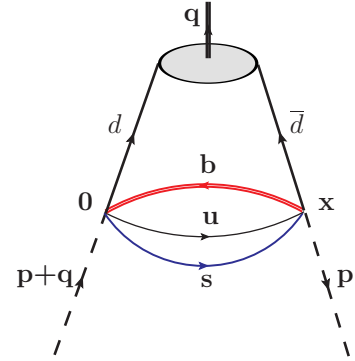}
\caption{The leading order diagram contributing to $\Pi _{1}(p,q)$. }
\label{fig:PertD}
\end{figure}

\begin{figure}[h]
\includegraphics[width=7cm]{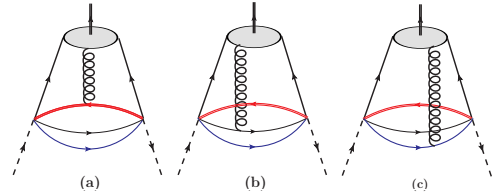}
\caption{The one-gluon exchange diagrams giving rise to corrections, which
can be computed by utilizing $\protect\rho $-meson three-particle DAs.}
\label{fig:Npert1}
\end{figure}

\begin{figure}[h]
\includegraphics[width=4.5cm]{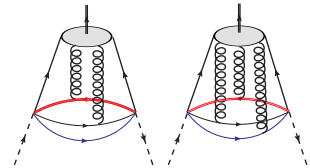}
\caption{Some many-particle diagrams neglected in this work.}
\label{fig:Npert2}
\end{figure}

The Feynman diagrams corresponding, for example, to the term $\Pi _{1}(p,q)$
are depicted in Figs.\ \ref{fig:PertD}, \ref{fig:Npert1} and \ref{fig:Npert2}.
The leading order contribution comes from the diagram shown in Fig.\ \ref{fig:PertD},
which corresponds to the term $\Pi _{1}^{\mathrm{pert}.}(p,q)$,
where all of the propagators are replaced by their perturbative components:
contribution of this diagram can be computed using the $\rho $-meson two
particle twist-two and higher twist distribution amplitudes. The diagrams
drawn in Fig.\ \ref{fig:Npert1} are obtained by choosing in one of the
propagators its $\sim G^{\mu \nu }$ component. They will be expressed in
terms of the meson's three-particle DAs. In this work we neglect corrections
arising from the diagrams (see, Fig.\ \ref{fig:Npert2} for some samples),
where in two or three propagators components $\sim G^{\mu \nu }$ are
chosen simultaneously. These contributions require invoking four and
five-particle distributions of the $\rho $-meson, and are beyond the scope
of the present study.

To provide some details of the calculations, as an example, we choose the
term $\Pi _{1}(p,q)$. The similar consideration can also be carried our for $%
\Pi _{2}(p,q)$. We start our analysis from the perturbative component of $%
\Pi _{1}(p,q)$ (Fig. \ref{fig:PertD}), i.e. from the contribution
\begin{eqnarray}
&&\Pi _{1}^{\mathrm{pert.}}(p,q)=i\int d^{4}xe^{ipx}\varepsilon
^{abc}\varepsilon ^{ade}\varepsilon ^{a^{\prime }b^{\prime }c^{\prime
}}\varepsilon ^{a^{\prime }d^{\prime }e^{\prime }}  \notag \\
&&\times \mathrm{Tr}\left[ \gamma _{5}\widetilde{S}_{s}^{b^{\prime }b(%
\mathrm{pert.})}(x)\gamma _{5}S_{u}^{cc^{\prime }(\mathrm{pert.})}(x)\right]
\notag \\
&&\times \left[ \gamma _{5}\widetilde{S}_{b}^{d^{\prime }d(\mathrm{pert.}%
)}(-x)\gamma _{5}\right] _{\alpha \beta }\langle \rho (q)|\overline{d}%
_{\alpha }^{e}(x)d_{\beta }^{e^{\prime }}(0)|0\rangle .  \label{eq:CF2pert}
\end{eqnarray}%
It is convenient first to perform the summation over the color indices. To
this end, we apply the projector onto the color singlet product of quarks
fields $\frac{1}{3}\delta _{ee^{\prime }}$ by performing the replacement
\begin{equation}
\overline{d}_{\alpha }^{e}(x)d_{\beta }^{e^{\prime }}(0)\rightarrow \frac{1}{%
3}\delta _{ee^{\prime }}\overline{d}_{\alpha }(x)d_{\beta }(0),
\end{equation}%
and use the expansion
\begin{equation}
\overline{d}_{\alpha }(x)d_{\beta }(0)\equiv \frac{1}{4}\Gamma _{\beta
\alpha }^{J}\overline{d}(x)\Gamma ^{J}d(0),  \label{eq:Expan}
\end{equation}%
where the sum runs over $J$
\begin{equation*}
\Gamma ^{J}=\mathbf{1,\ }\gamma _{5},\ \gamma _{\mu },\ i\gamma _{5}\gamma
_{\mu },\ \sigma _{\mu \nu }/\sqrt{2}.
\end{equation*}%
Substituting this expansion into Eq.\ (\ref{eq:CF2pert}) we obtain
\begin{eqnarray}
&&\Pi _{1}^{\mathrm{pert.}}(p,q)=i\int d^{4}xe^{ipx}\mathrm{Tr}\left[ \gamma
_{5}\widetilde{S}_{s}^{(\mathrm{pert.})}(x)\gamma _{5}S_{u}^{(\mathrm{pert.}%
)}(x)\right]  \notag \\
&&\times \mathrm{Tr}\left[ \gamma _{5}\widetilde{S}_{b}^{(\mathrm{pert.}%
)}(-x)\gamma _{5}\Gamma ^{J}\right] \langle \rho (q)|\overline{d}(x)\Gamma
^{J}d(0)|0\rangle .  \label{eq:CFpert}
\end{eqnarray}%
Now, as an example, we analyze the nonperturbative diagram depicted in
Fig. \ref{fig:Npert1}(b). After some manipulations we recast the
corresponding function $\Pi_{1(b)}^{\mathrm{n.-pert.}}(p,q)$ into the form
\begin{eqnarray}
&&\Pi _{1(b)}^{\mathrm{n.-pert.}}(p,q)=i\int d^{4}xe^{ipx}\mathrm{Tr}\left[
\gamma _{5}\widetilde{S}_{s}^{(\mathrm{pert.})}(x)\gamma _{5}\right.  \notag
\\
&&\left. \left\{ -ig_{s}\int_{0}^{1}dv\frac{1}{16\pi ^{2}x^{2}}\left[ {%
\slashed x}\sigma ^{\mu \nu }-4ivx^{\mu }\gamma ^{\nu }\right] \right\} %
\right]  \notag \\
&&\times \mathrm{Tr}\left[ \gamma _{5}\widetilde{S}_{b}^{(\mathrm{pert.}%
)}(-x)\gamma _{5}\Gamma ^{J}\right] \frac{1}{4}\langle \rho (q)|\overline{d}%
(x)\Gamma ^{J}G_{\mu \nu }(vx)d(0)|0\rangle .  \notag \\
&&{}  \label{eq:NPT}
\end{eqnarray}%
The similar analysis can be done for other nonperturbative diagrams, as well.

The sum of the $\Pi _{1}^{\mathrm{pert.}}(p,q)$ and $\Pi _{1(i)}^{\mathrm{%
n.-pert.}}(p,q)$ for $i=a,\ b$ and $c$ determines the first component $\Pi
_{1}(p,q)$ of the correlation function. It is given as the integral of the
products of the coefficient functions and non-local matrix elements
\begin{eqnarray}
&&\langle \rho (q)|\overline{d}(x)\Gamma ^{J}d(0)|0\rangle ,  \notag \\
&&\langle \rho (q)|\overline{d}(x)\Gamma ^{J}G_{\mu \nu }(vx)d(0)|0\rangle .
\label{eq:MatElem}
\end{eqnarray}%
The matrix elements of the neutral $\rho$ meson from Eq.\ (\ref{eq:MatElem})
up to an isospin factor in the overall normalization are connected with ones
of the charged $\rho$ mesons, and can be expanded in terms of the corresponding distribution
amplitudes. Below we provide expressions for the $\langle 0|\overline{%
u}(x)\Gamma ^{J}d(0)|\rho (q)\rangle $ type matrix elements obtained to twist-4
accuracy and given by means of the $\rho $ meson's two-particle DAs. For the
structures $\Gamma ^{J}=\mathbf{1}$ and $\mathbf{\ }\gamma _{\mu }\gamma
_{5} $ we get%
\begin{eqnarray}
&&\langle 0|\overline{u}(x)d(0)|\rho (q)\rangle =-if_{\rho }^{\perp
}\varepsilon \cdot xm_{\rho }^{2}\int_{0}^{1}due^{i\overline{u}qx}\psi
_{3}^{\parallel }(u),  \notag \\
&&\langle 0|\overline{u}(x)\gamma _{\mu }\gamma _{5}d(0)|\rho (q)\rangle =%
\frac{1}{2}f_{\rho }^{\parallel }m_{\rho }\epsilon _{\mu }^{\nu \alpha \beta
}\varepsilon _{\nu }q_{\alpha }x_{\beta }  \notag \\
&&\times \int_{0}^{1}due^{i\overline{u}qx}\psi _{3}^{\perp }(u),
\end{eqnarray}%
whereas $\Gamma ^{J}=\gamma _{\mu }\ $and $\sigma _{\mu \nu }$ give
\begin{eqnarray}
&&\langle 0|\overline{u}(x)\gamma _{\mu }d(0)|\rho (q)\rangle =f_{\rho
}^{\parallel }m_{\rho }\left\{ \frac{\varepsilon \cdot x}{q\cdot x}q_{\mu
}\right.  \notag \\
&&\times \int_{0}^{1}due^{i\overline{u}qx}\left[ \phi _{2}^{\parallel }(u)+%
\frac{m_{\rho }^{2}x^{2}}{4}\phi _{4}^{\parallel }(u)\right]  \notag \\
&&+\left( \varepsilon _{\mu }-q_{\mu }\frac{\varepsilon \cdot x}{q\cdot x}%
\right) \int_{0}^{1}due^{i\overline{u}qx}\phi _{3}^{\perp }(u)  \notag \\
&&\left. -\frac{1}{2}x_{\mu }\frac{\varepsilon \cdot x}{(q\cdot x)^{2}}%
m_{\rho }^{2}\int_{0}^{1}due^{i\overline{u}qx}C(u)+...\right\} ,
\label{eq:DAT2}
\end{eqnarray}%
and
\begin{eqnarray}
&&\langle 0|\overline{u}(x)\sigma _{\mu \nu }d(0)|\rho (q)\rangle =if_{\rho
}^{\perp }\left\{ \left( \varepsilon _{\mu }q_{\nu }-\varepsilon _{\nu
}q_{\mu }\right) \right.  \notag \\
&&\times \int_{0}^{1}due^{i\overline{u}qx}\left[ \phi _{2}^{\perp }(u)+\frac{%
m_{\rho }^{2}x^{2}}{4}\phi _{4}^{\perp }(u)\right]  \notag \\
&&+\frac{1}{2}\left( \varepsilon _{\mu }x_{\nu }-\varepsilon _{\nu }x_{\mu
}\right) \frac{m_{\rho }^{2}}{q\cdot x}\int_{0}^{1}due^{i\overline{u}qx}%
\left[ \psi _{4}^{\perp }(u)-\phi _{2}^{\perp }(u)\right]  \notag \\
&&\left. +\left( q_{\mu }x_{\nu }-q_{\nu }x_{\mu }\right) \frac{\varepsilon
\cdot x}{(q\cdot x)^{2}}m_{\rho }^{2}\int_{0}^{1}due^{i\overline{u}%
qx}D(u)+...\right\}  \notag \\
&&{}  \label{eq:DAT3}
\end{eqnarray}%
respectively. Here $\overline{u}=1-u$, and $m_{\rho }$ , $\varepsilon $ are
the mass of $\rho $ meson and its polarization vector. In the equations
above the functions $C(u)$ and $D(u)$ denote the following combinations of
the two-particle DAs
\begin{eqnarray}
&&C(u)=\psi _{4}^{\parallel }(u)+\phi _{2}^{\parallel }(u)-2\phi _{3}^{\perp
}(u), \\
&&D(u)=\phi _{3}^{\parallel }(u)-\frac{1}{2}\phi _{2}^{\perp }(u)-\frac{1}{2}%
\psi _{4}^{\perp }(u).
\end{eqnarray}%
The twists of the distribution amplitudes are shown as subscripts in the
relevant functions. As is seen, these matrix elements include the
two-particle leading twist DAs $\phi _{2}^{\parallel (\perp )}(u)$, the
twist-3 distribution amplitudes $\phi _{3}^{\parallel (\perp )}(u)$ and $%
\psi _{3}^{\parallel (\perp )}(u)$, as well as twist-4 distributions $\phi
_{4}^{\parallel (\perp )}(u)$ and $\psi _{4}^{\parallel (\perp )}(u)$.

We do not write down here lengthy equalities, which express the matrix
elements $\langle \rho (q)|\overline{d}(x)\Gamma ^{J}G_{\mu
\nu}(vx)d(0)|0\rangle$ in terms of the numerous higher twist DAs of the $%
\rho $ meson, and refrain from giving further information on the DAs
themselves. The definitions and detailed information on
properties of the distribution amplitudes of the $\rho$ and other vector mesons,
as well as explicit expressions for some of their models, used also in the present work,
can be found in Refs.\ \cite{Ball:1996tb,Ball:1998sk,Ball:1998ff,Ball:2007rt,Ball:2007zt}.

Our aim is to calculate the correlation function $\Pi ^{\mathrm{QCD}}(p,q)$
in terms of the DAs of the $\rho $ meson, extract the invariant amplitude $%
\Pi ^{\mathrm{QCD}}(p^{2},p^{\prime 2})$ corresponding to the structure $%
p\cdot \varepsilon ,$ and perform its double Borel transformation
\begin{equation*}
\Pi ^{\mathrm{QCD}}(M^{2})=\mathcal{B}_{p^{2}}(M_{1}^{2})\mathcal{B}%
_{p^{\prime 2}}(M_{2}^{2})\Pi ^{\mathrm{QCD}}(p^{2},p^{\prime 2}).
\end{equation*}%
After equating $\Pi ^{\mathrm{QCD}}(M^{2})$ to its counterpart $\Pi ^{%
\mathrm{Phys}}(M^{2})$ and subtracting  contributions of the higher resonances and
continuum states presented in Eq.\ (\ref{eq:BorT}) as the double dispersion integral,
we can derive the LCSR for the strong coupling $G_{X_{b}X_{b}\rho }$.

Presenting some details of calculations in Appendix \ref{sec:App1}, below we write down the final expression obtained
for $\Pi _{1}^{\mathrm{QCD}}(M^{2})$
\begin{eqnarray}
\Pi _{1}^{\mathrm{QCD}}(M^{2}) &=&\frac{m_{b}m_{\rho }}{64\pi ^{4}}%
\int_{m_{b}^{2}}^{\infty }dse^{(m_{\rho }^{2}-4s)/4M^{2}}\left[ \Gamma
\left( M^{8},s\right) \right.   \notag \\
&&\left. +\Gamma \left( M^{6},s\right) +\Gamma \left( M^{4},s\right) \right]
.
\end{eqnarray}%
Here
\begin{equation}
\Gamma \left( M^{8},s\right) =-2m_{b}^{3}f_{\rho }^{\parallel }M^{8}\phi
_{2}^{\parallel }(\overline{u}_{0})\left( \frac{1}{s^{3}}-\frac{2m_{b}^{2}}{%
s^{4}}+\frac{m_{b}^{4}}{s^{5}}\right) ,
\end{equation}%
\begin{eqnarray}
&&\Gamma \left( M^{6},s\right) =-m_{b}m_{\rho }M^{6}\left\{ m_{b}^{2}m_{\rho
}f_{\rho }^{\parallel }\left( \frac{m_{b}^{2}}{s^{4}}-\frac{1}{s^{3}}\right)
\right.   \notag \\
&&\times \phi _{4}^{\parallel }(\overline{u}_{0})+m_{\rho }f_{\rho
}^{\parallel }\left. \Bigg[\left( \frac{1}{s^{2}}-\frac{2m_{b}^{2}}{s^{3}}+%
\frac{m_{b}^{4}}{s^{4}}\right) \right.   \notag \\
&&\times \left[ \mathit{I}_{1}\left( \widetilde{\Phi }_{3}^{\parallel
}(\alpha ),1\right) -3\mathit{I}_{1}\left( \Phi _{4}^{\parallel }(\alpha
),1\right) +6\mathit{I}_{1}\left( \Phi _{4}^{\parallel }(\alpha ),v\right)
\right.   \notag \\
&&+3\mathit{I}_{1}\left( \widetilde{\Phi }_{4}^{\parallel }(\alpha
),1\right) -\mathit{I}_{1}\left( \Psi _{4}^{\parallel }(\alpha ),1\right) +2%
\mathit{I}_{1}\left( \Psi _{4}^{\parallel }(\alpha ),v\right)   \notag \\
&&\left. +\mathit{I}_{1}\left( \widetilde{\Psi }_{4}^{\parallel }(\alpha
),1\right) -\mathit{I}_{1}\left( \Phi _{3}^{\parallel }(\alpha ),1\right) +2%
\mathit{I}_{1}\left( \Phi _{3}^{\parallel }(\alpha ),v\right) \right]
\notag \\
&&\left. +8m_{b}^{2}\left( \frac{m_{b}^{2}}{s^{4}}-\frac{1}{s^{3}}\right)
\mathit{I}_{2}[C(u_{0})]\right. \Bigg ]  \notag \\
&&\left. +4m_{b}f_{\rho }^{\perp }\left( \frac{1}{s^{2}}-\frac{2m_{b}^{2}}{%
s^{3}}+\frac{m_{b}^{4}}{s^{4}}\right) \psi _{3}^{\parallel }(\overline{u}%
_{0})\right\}   \label{eq:M6}
\end{eqnarray}%
and
\begin{eqnarray}
&&\Gamma \left( M^{4},s\right) =m_{\rho }^{3}M^{4}\left\{ f_{\rho }^{\perp
}\left( \frac{1}{s}-\frac{2m_{b}^{2}}{s^{2}}+\frac{m_{b}^{4}}{s^{3}}\right)
\right.   \notag \\
&&\left[ 3\mathit{I}_{0}\left( \Phi _{3}^{\perp }(\alpha ),1\right) -2\left(
\mathit{I}_{0}\left( \Phi _{4}^{\perp (3)}(\alpha ),1\right) \right. \right.
\notag \\
&&\left. \left. +\mathit{I}_{0}\left( \Phi _{4}^{\perp (4)}(\alpha
),1\right) \right) \right] +8m_{b}m_{\rho }f_{\rho }^{\parallel }\left(
\frac{1}{s^{2}}-\frac{m_{b}^{2}}{s^{3}}\right)   \notag \\
&&\left. \times \mathit{I}_{0}\left( \Psi _{4}^{\parallel }(\alpha
),k-u_{0}\right) \right\}.
\label{eq:M4}
\end{eqnarray}%
In the formulas presented above, we introduce shorthand notations for some
integrals. Namely, we use
\begin{eqnarray}
\mathit{I}_{0}\left( \Phi (\alpha ),k-u_{0}\right)  &=&\int \mathcal{D}%
\alpha \int_{0}^{1}dv\left( k-u_{0}\right) \Phi (\alpha _{\overline{q}%
},\alpha _{q},\alpha _{g})  \notag \\
&&\times \theta \left( k-u_{0}\right) ,  \label{eq:I0}
\end{eqnarray}%
\begin{eqnarray}
\mathit{I}_{1}\left( \Phi (\alpha ),f(v)\right)  &=&\int \mathcal{D}\alpha
\int_{0}^{1}dv\Phi (\alpha _{\overline{q}},\alpha _{q},\alpha _{g})  \notag
\\
&&\times f(v)\delta \left( k-u_{0}\right) ,  \label{eq:I1}
\end{eqnarray}%
and
\begin{equation}
\mathit{I}_{2}\left( C(u_{0})\right) =\int_{0}^{1-u_{0}}du^{\prime }\left(
u_{0}+u^{\prime }-1\right) C\left( u^{\prime }\right) .  \label{eq:I2}
\end{equation}%
In Eqs.\ (\ref{eq:I0}), (\ref{eq:I1}) and (\ref{eq:I2})
\begin{equation*}
k=\alpha _{\overline{q}}+\alpha _{g}(1-v),
\end{equation*}%
and the integration measure $\mathcal{D}\alpha $ is defined as
\begin{equation*}
\int \mathcal{D}\alpha =\int_{0}^{1}d\alpha _{q}\int_{0}^{1}d\alpha _{\bar{q}%
}\int_{0}^{1}d\alpha _{g}\delta (1-\alpha _{q}-\alpha _{\bar{q}}-\alpha
_{g}).
\end{equation*}
The similar calculations have been carried out to derive the second
component of the correlation function $\Pi_2^{\mathrm{QCD}}(M^2)$.

As we have noted above, the sum rules for the coupling $G_{X_bX_b\rho}$ can be derived
after continuum subtraction. The contribution coming from the higher resonances and continuum
states is written down in Eq.\ (\ref{eq:BorT}) as the double dispersion integral over the
physical spectral density $\rho^{\mathrm{phys}}(s_1,s_2)$. The subtraction is performed invoking
ideas of the quark-hadron duality, i.e. by assuming that in some regions of physical quantities $\rho^{\mathrm{phys}}(s_1,s_2)$ may be replaced by its theoretical counterpart
$\rho^{\mathrm{QCD}}(s_1,s_2)$, the latter is being calculable within the perturbative QCD.
The spectral density $\rho^{\mathrm{QCD}}(s_1,s_2)$ may be found by computing the imaginary part
of the correlation function, or extracted directly from its Borel transformed expression using a
technique, which is described in Refs.\ \cite{Belyaev:1994zk,Aliev:2010yx,Aliev:2011ufa,Ball:1994}.
Then the continuum subtraction can be performed in accordance with the prescriptions developed in these papers.
It is based on the observation that double spectral density of the leading contributions $\sim M^{2}$, is
concentrated  near the diagonal $s_1=s_2$. In this case for the continuum
subtraction the simple expressions can be derived, which are not sensitive to the shape of the
duality region. In the case $M_{1}^{2}=M_{2}^{2}=2M^{2}$ and $u_{0}=1/2$, for example, the factor
\begin{equation}
\left( M^{2}\right) ^{N}e^{-m^{2}/M^{2}}
\end{equation}%
remains in its original form if $N\leq 0$,
and is replaced as
\begin{equation}
\left( M^{2}\right) ^{N}e^{-m^{2}/M^{2}} \to
\frac{1}{\Gamma (N)}\int_{m^{2}}^{s_0}dse^{-s/M^{2}}\left( s-m^{2}\right)
^{N-1},
\end{equation}%
for $N>0$. The subtracted version of other expressions, which may encounter in the sum rule calculations are collected in Appendix \ref{sec:App2}. In the present work we follow these procedures to perform the continuum subtraction.

\section{Numerical results}
\label{sec:Num}

The sum rules for the strong couplings contain some
parameters, which should be determined to carry out the numerical
computations. The mass and current coupling of the exotic $X_b$ state, as well
as the mass and decay constants of the $\rho$ meson are among the important
physical parameters of the problem under consideration. The situation with the
$\rho$ meson is clear, because its parameters are well known: they were
extracted from experimental data or evaluated employing various
nonperturbative approaches, including the LCSR method \cite{Ball:2007zt,Agashe:2014kda}. The relevant
information is given in Table \ref{tab:Param}.

The parameters of the $X_b$ state deserve more detailed consideration. Thus,
its mass $m_{X_b}$, decay constant $f_{X_b}$ and the width of the decay $X_b
\to B_s \pi$ were calculated in our previous works (see, \cite%
{Agaev:2016mjb,Agaev:2016ijz}) using a vector diquark-vector antidiquark type
interpolating current. The same parameters were also computed in Ref.\ \cite{Agaev:2016urs} by suggesting
the molecule-type internal structure for the $X_b$ state.
\begin{table}[t]
\begin{tabular}{|c|c|}
\hline\hline
Parameters & Values \\ \hline\hline
$m_{\rho }$ & $(775.26 \pm 0.25)~\mathrm{MeV}$ \\
$f_{\rho }^{\parallel }$ & $(0.216\pm 0.003)~\mathrm{GeV}$ \\
$f_{\rho }^{\perp }$ & $(0.165\pm 0.009)~\mathrm{GeV} $ \\
$a_{2}^{\parallel }$ & $0.15\pm 0.07 $ \\
$a_{2}^{\perp } $ & $0.14\pm 0.06$ \\ \hline\hline
\end{tabular}%
\caption{The mass, decay constants, and parameters of the $\protect\rho$
meson leading twist DAs.}
\label{tab:Param}
\end{table}

In the present study, as an intermediate stage of the full analysis, we would like
to calculate the spectroscopic parameters of the $X_b$ state using the interpolating current adopted in the
present work (see, Eq.\ (\ref{eq:CR1})). Our predictions for the mass
\begin{equation}
m_{X_b}=(5620 \pm 195 )\, \mathrm{MeV},  \label{eq:MassN}
\end{equation}
found by this way, is slightly larger than one given in Ref.\ \cite%
{Agaev:2016mjb}, but still in agreement with the data of
the D0 Collaboration. For the current coupling  $f_{X_b}$  we obtain
\begin{equation}
f_{X_b}=(0.14 \pm 0.02)\times 10^{-2}\, \mathrm{GeV}^{4}.  \label{eq:DCNew}
\end{equation}

We utilize the masses of the heavy quarks in the $\overline{MS}$ scheme
\begin{eqnarray}
&&m_b(m_b)=4.18 \pm 0.03\, \mathrm{GeV},  \notag \\
&&m_c(m_c)=1.275 \pm 0.025\, \mathrm{GeV}.
\end{eqnarray}
The scale dependence of $m_b$ and $m_c$ is taken into account in accordance
with the renormalization group evolution
\begin{equation}
m_{q}(\mu)=m_{q}(\mu_0)\left [ \frac{\alpha_s(\mu)}{\alpha_s(\mu_0)}\right
]^{\gamma_{q}},
\label{eq:Evol}
\end{equation}
with $\gamma_b=12/23$ and $\gamma_c=12/25$. The renormalization scale in
computation of the coupling $G_{X_bX_b\rho}$ is taken equal to
\begin{equation}
\mu_b=\sqrt {m_{X_{b}}^2-(m_{b}+m_s)^{2}}\simeq 3.598\,\,\mathrm{GeV}.
\end{equation}
The mass of the $b$ quark is evolved to this scale by employing the two-loop
QCD running coupling $\alpha_{s}(\mu)$ with $\Lambda^{(4)}=326\,\,\mathrm{MeV%
}$.

Another set of parameters is formed due to various distribution amplitudes
of the $\rho$ meson. Indeed, the leading and higher twist DAs are the
important ingredients of the LCSR expressions, and in turn, contain numerous
parameters. The leading twist DAs of the longitudinally and transversely polarized
$\rho$ meson are given by the formula
\begin{equation}
\phi _{2}^{\parallel (\perp) }(u)=6u\overline{u}\left[ 1+\sum_{n=2}^{%
\infty}a_{n}^{\parallel (\perp) }C_{n}^{3/2}(2u-1)\right],  \label{eq:LTDA}
\end{equation}
where $C_{n}^{m}(z)$ are the Gegenbauer polynomials. Equation (\ref{eq:LTDA}%
) is the general expression for $\phi _{2}^{\parallel (\perp) }(u)$. In our
calculations we employ twist-2 DAs with only one non-asymptotic term, i.e.
only the coefficients $a_{2}^{\parallel (\perp) } \neq 0$ (see, Table \ref%
{tab:Param}).
The models for the higher twist DAs, which enter to Eqs.\ (\ref%
{eq:M6}) and (\ref{eq:M4}) are borrowed from Refs.\ \cite%
{Ball:2007rt,Ball:2007zt}. The values of the relevant parameters at the
normalization scale $\mu_0=1\, \mathrm{GeV}$ can be found in Tables 1 and 2
of Ref.\ \cite{Ball:2007zt}.

\begin{figure}[tb]
\includegraphics[width=8.cm]{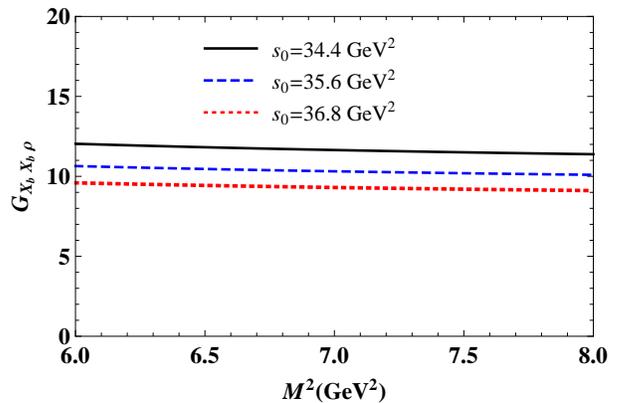}
\caption{The strong coupling $G_{X_bX_b\protect\rho}$ as a function of the
Borel parameter $M^2$ at different values of $s_0$.}
\label{fig:Coupl1}
\end{figure}

Finally, the sum rule expressions depend on two auxiliary parameters, i.e. on the
Borel parameter $M^2$ and continuum threshold $s_0$, which are
unavoidable within this method. Results, in general, should not depend on the choice of $M^2$ and $%
s_0$. In practice, however, one may only minimize effects connected with
their variations. Exploring the obtained sum rules we fix
working windows within of which the parameters $s_0$ and $M^2$ can be
varied: for the threshold $s_0$ we find
\begin{equation}
34.4\,\,\mathrm{GeV}^2 \leq s_0 \leq 36.8\,\, \mathrm{GeV}^2,
\end{equation}
whereas the Borel parameter can be varied in the limits
\begin{equation}
6\,\,\mathrm{GeV}^2 \leq M^{2} \leq 8\,\, \mathrm{GeV}^2.
\end{equation}

The results of computations are depicted in Fig. \ref{fig:Coupl1}. In
accordance with our studies, the strong coupling $G_{X_bX_b\rho}$ is equal
to
\begin{equation}
G_{X_bX_b\rho}=10.46 \pm 2.26.
\end{equation}

The similar analysis in the case of the vertex $X_{c}X_{c}\rho $ using the parameters of the $X_c$ state, namely
\begin{eqnarray}
&&m_{X_c}=(2634 \pm 62 )\, \mathrm{MeV},  \notag \\
&&f_{X_c}=(0.11 \pm 0.02)\times 10^{-2}\, \mathrm{GeV}^{4},
\label{eq:ParXc}
\end{eqnarray}
given in Ref.\ \cite{Agaev:2016lkl}, restricts $s_{0}$ and $M^{2}$ inside the ranges:
\begin{eqnarray}
7.6\,\mathrm{GeV}^{2} &\leq &s_{0}\leq 8.1\,\mathrm{GeV}^{2}, \\
3\,\mathrm{GeV}^{2} &\leq &M^{2}\leq 5\,\mathrm{GeV}^{2}.
\end{eqnarray}%
The scale dependence of $m_c$ is taken into account in accordance
with Eq.\ (\ref{eq:Evol}), where
\begin{equation}
\mu_c=\sqrt {m_{X_{c}}^2-(m_{c}+m_s)^{2}}\simeq 2.224\,\,\mathrm{GeV}.
\end{equation}
As in the previous case, the mass of the $c$ quark is evolved to the scale $\mu_c$ by employing the two-loop
QCD running coupling $\alpha_{s}(\mu)$.

\begin{figure}[h]
\includegraphics[width=8.2cm]{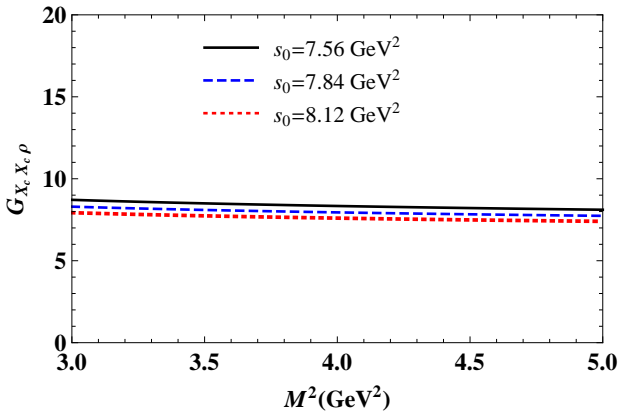}
\caption{The coupling $G_{X_cX_c\protect\rho}$ vs the Borel parameter $M^2$
at different values of $s_0$.}
\label{fig:Coupl2}
\end{figure}

The results of the numerical calculations are shown in Fig.\ \ref{fig:Coupl2}. The QCD light-cone sum rule prediction for the strong coupling $G_{X_{c}X_{c}\rho }$ extracted in the present work reads:
\begin{equation}
G_{X_{c}X_{c}\rho }=8.01\pm 1.66.
\end{equation}

In the present work, we applied for the first time the full theory of the QCD
light-cone sum rule method to systems of the tetraquarks with  a single heavy quark,
and calculated the strong couplings of the $X_b$ and $X_c$ states with the $\rho$ meson.
To this end, we derived the sum rules by equating the Borel transformations of the same
correlation function found in terms of physical quantities to its expression obtained
by employing the leading and higher twist distribution amplitudes of the $\rho$ meson.
We also demonstrated that technical tools elaborated for analysis of the transition
form factors and strong couplings of the conventional hadrons, in general, are applicable
to these complicated quark systems, as well.

\section*{ACKNOWLEDGEMENTS}
The work of S.~S.~A. was supported by the TUBITAK grant 2221-"Fellowship
Program For Visiting Scientists and Scientists on Sabbatical Leave". This
work was also supported in part by TUBITAK under the grant no: 115F183.

\appendix
\section{ Calculation of $\Pi _{1}^{\mathrm{QCD}}(M^2)$: some details}
\renewcommand{\theequation}{\Alph{section}.\arabic{equation}}
\label{sec:App1}

In this Appendix  we provide some details of the calculations of the function
$\Pi _{1}^{\mathrm{QCD}}(M^2)$. To this end, we pick up a simple term from the
perturbative component given by Eq.\ (\ref%
{eq:CFpert}) and a term $\sim \phi (u)$ from the expression of the
distribution amplitude. Obtained by this way integral has the form
\begin{equation}
I=\int_{0}^{1}du\phi (u)\int d^{4}xe^{ipx+i\overline{u}qx}\frac{1}{x^{2n}}%
\frac{K_{\nu }\left( m_{Q}\sqrt{-x^{2}}\right) }{\left( \sqrt{-x^{2}}\right)
^{\upsilon }}.  \label{eq:Int}
\end{equation}%
In Eq.\ (\ref{eq:Int}) the factor $1/x^{2(n_{1}+n_{2})}\equiv $ $1/x^{2n}$
is due to the light quark propagators, whereas the factor $\sim $ $K_{\nu }$
comes from the heavy quark propagator. To proceed we apply the integral
representation for the Bessel functions
\begin{equation*}
\frac{K_{\nu }\left( m_{Q}\sqrt{-x^{2}}\right) }{\left( \sqrt{-x^{2}}\right)
^{\upsilon }}=\frac{1}{2}\int_{0}^{\infty }\frac{dt}{t^{\nu +1}}\exp \left[ -%
\frac{m_{Q}}{2}\left( t-\frac{x^{2}}{t}\right) \right] ,
\end{equation*}%
and perform the Wick rotation, i.e. replace $x^{2}=-\widetilde{x}^{2}$, $%
px\rightarrow -\widetilde{p}\widetilde{x}$, and $qx\rightarrow -\widetilde{q}%
\widetilde{x}$. \ Finally, we make use of the Schwinger representation for
the terms $1/\widetilde{x}^{2n}$
\begin{equation*}
\frac{1}{(\widetilde{x}^{2})^{n}}=\frac{1}{\Gamma (n)}\int_{0}^{\infty
}d\lambda \lambda ^{n-1}\exp \left( -\lambda \widetilde{x}^{2}\right) ,
\end{equation*}%
and, in what follows, omit the tilde on these variables. These replacements
yield
\begin{eqnarray}
I &=&\frac{i}{\Gamma (n)}\int_{0}^{1}du\phi (u)\int_{0}^{\infty }\frac{dt}{%
t^{\nu +1}}\exp \left[ -\frac{m_{Q}}{2}t\right] \int_{0}^{\infty }d\lambda
\lambda ^{n-1}  \notag \\
&&\times \int d^{4}x\exp \left[ -ipx-i\overline{u}qx-\lambda x^{2}-\frac{%
m_{Q}}{2}\frac{x^{2}}{t}\right] .  \label{eq:IntE}
\end{eqnarray}%
Having shifted the variable $x$ as%
\begin{equation*}
x\rightarrow x-\frac{i(p+\overline{u}q)}{2\left( \lambda +m_{Q}/2t\right) },
\end{equation*}%
and performed the four-dimensional Gaussian integral over the new $x$ we find%
\begin{eqnarray*}
&&\int d^{4}x\exp \left[ -ipx-i\overline{u}qx-\lambda x^{2}-\frac{m_{Q}}{2}%
\frac{x^{2}}{t}\right]  \\
&=&\left( \frac{2\pi t}{m_{Q}+2\lambda t}\right) ^{2}\exp \left[ -\frac{t(p+%
\overline{u}q)^{2}}{2(m_{Q}+2\lambda t)}\right] .
\end{eqnarray*}%
The Borel transformations of the integral $I$ give
\begin{eqnarray}
&&I\sim \frac{i}{\Gamma (n)}\int_{0}^{1}du\phi (u)\int_{0}^{\infty }\frac{dt%
}{t^{\nu +1}}e^{-\frac{m_{Q}}{2}t}\int_{0}^{\infty }d\lambda \lambda ^{n-1}
\notag \\
&&\times \exp \left[ \frac{tu\overline{u}}{2(m_{Q}+2\lambda t)}q^{2}\right]
\delta \left( \frac{1}{M_{1}^{2}}-\frac{tu}{2(m_{Q}+2\lambda t)}\right)
\notag \\
&&\times \delta \left( \frac{1}{M_{2}^{2}}-\frac{t\overline{u}}{%
2(m_{Q}+2\lambda t)}\right) .
\end{eqnarray}%
Now using
\begin{equation}
\delta \left( \frac{1}{M_{1}^{2}}-\frac{tu}{2(m_{Q}+2\lambda t)}\right) =%
\frac{M_{1}^{4}u}{4}\delta (\lambda -\lambda _{0})\theta (\lambda _{0}),
\end{equation}%
where $\lambda _{0}$ equals to
\begin{equation}
\frac{M_{1}^{2}tu-2m_{Q}}{4t},
\end{equation}%
we carry out the $\lambda $ integration. The next step is computation of the
$u$ integral. To this end, we employ the second delta function and transform
it as
\begin{equation*}
\delta \left( \frac{1}{M_{2}^{2}}-\frac{t\overline{u}}{2(m_{Q}+2\lambda
_{0}t)}\right) =\frac{M_{1}^{2}M_{2}^{4}}{\left( M_{1}^{2}+M_{2}^{2}\right)
^{2}}\delta (u-u_{0}),
\end{equation*}%
where%
\begin{equation*}
u_{0}=\frac{M_{2}^{2}}{M_{1}^{2}+M_{2}^{2}}.
\end{equation*}%
The integration over $u$ sets $\phi (u)\rightarrow \phi (u_{0})$, and also
determines the low limit of the remaining $t$ integral, which has become
equal to
\begin{equation*}
t_{\min }=\frac{2m_{Q}}{M_{1}^{2}u_{0}}.
\end{equation*}%
By re-scaling the variable $t$
\begin{equation*}
t\rightarrow \frac{2}{M_{1}^{2}u_{0}}\frac{s}{m_{Q}},
\end{equation*}%
we obtain the integral over $s$ running from $m_{Q}^{2}$ till infinity, and,
by this way, the considering component of $\Pi _{1}^{\mathrm{QCD}}(M^{2})$ takes
its final form.

\section{The formulas for the continuum subtraction}
\renewcommand{\theequation}{\Alph{section}.\arabic{equation}}
\label{sec:App2}

Here we have collected useful formulas, which can be applied in the continuum
subtraction. In the left-hand side of the formulas presented below we write
down the original form, and in the right-hand side the subtracted version of
expressions encountered in the sum rule calculations:
\begin{equation}
\left( M^{2}\right) ^{N}\int_{m^{2}}^{\infty }dse^{-s/M^{2}}f(s)\rightarrow
\int_{m^{2}}^{s_{0}}dse^{-s/M^{2}}F_{N}(s).
\end{equation}%
For the more complicated factor
\begin{equation}
\left( M^{2}\right) ^{N}\ln \left( \frac{M^{2}}{\Lambda ^{2}}\right)
\int_{m^{2}}^{\infty }dse^{-s/M^{2}}f(s),
\end{equation}%
for all values of $N$ the following formula is valid
\begin{eqnarray}
&& \int_{m^{2}}^{s_{0}}dse^{-s/M^{2}}\left[ F_{N}(m^{2})\ln \left( \frac{%
s-m^{2}}{\Lambda ^{2}}\right) +\gamma _{E}F_{N}(s) \right. \notag \\
&&\left. +\int_{m^{2}}^{s}duF_{N-1}(u)\ln \left( \frac{s-u}{%
\Lambda ^{2}}\right) \right] .
\end{eqnarray}%

The next formula is
\begin{eqnarray}
&&\left( M^{2}\right) ^{N}\ln \left( \frac{M^{2}}{\Lambda ^{2}}\right)
e^{-m^{2}/M^{2}} \notag \\
&&\rightarrow e^{-s_{0}/M^{2}}\sum_{i=1}^{1-N}\left( \frac{d}{%
ds_{0}}\right) ^{1-N-i}\left[ \ln \left( \frac{s_{0}-m^{2}}{\Lambda ^{2}}%
\right) \right] \frac{1}{\left( M^{2}\right) ^{i-1}}  \notag \\
&&+\gamma _{E}\left( M^{2}\right) ^{N}\left( e^{-m^{2}/M^{2}}-\delta_{N1}e^{-s_{0}/M^{2}}\right) \notag \\
&&+\left( M^{2}\right)^{N-1}\int_{m^{2}}^{s_{0}}dse^{-s/M^{2}}\ln \left( \frac{s-m^{2}}{\Lambda
^{2}}\right),
\end{eqnarray}%
if $N\leq 1$, and
\begin{eqnarray}
&&\frac{\gamma _{E}}{\Gamma (N)}\int_{m^{2}}^{s_{0}}dse^{-s/M^{2}}\left(
s-m^{2}\right) ^{N-1} \notag \\
&&+\frac{1}{\Gamma (N-1)}%
\int_{m^{2}}^{s_{0}}dse^{-s/M^{2}}\int_{m^{2}}^{s}du(s-u)^{N-2} \notag \\
&&\times \ln \left(
\frac{u-m^{2}}{\Lambda ^{2}}\right),
\end{eqnarray}%
for  $N>1$.

Useful are also the expressions
\begin{eqnarray}
&&\left( M^{2}\right) ^{N}\int_{m^{2}}^{\infty }dse^{-s/M^{2}}f(s)\ln \left(
\frac{s-m^{2}}{\Lambda ^{2}}\right) \notag \\
&&\rightarrow e^{-s_{0}/M^{2}}\sum_{i=1}^{|N|}\frac{\widetilde{F}_{N+i}(s_{0})}{\left(
M^{2}\right) ^{i-1}}+\left( M^{2}\right)
^{N}\int_{m^{2}}^{s_{0}}dse^{-s/M^{2}}f(s) \notag \\
&&\times \ln \left( \frac{s-m^{2}}{\Lambda
^{2}}\right) ,\ \ \ N\leq 0,
\end{eqnarray}%
and%
\begin{eqnarray}
&&\frac{1}{\Gamma (N)}\int_{m^{2}}^{s_{0}}dse^{-s/M^{2}}%
\int_{m^{2}}^{s}du(s-u)^{N-1} \notag \\
&&\times \ln \left( \frac{u-m^{2}}{\Lambda ^{2}}\right)f(u)
,\ \ N>0.
\end{eqnarray}%
In the equations above we have employed the notations%
\begin{equation}
F_{N}(s)=\left( \frac{d}{ds}\right)^{-N}f(s), \,\,\, N\leq 0,
\end{equation}%
and%
\begin{equation}
F_{N}(s)=\frac{1}{\Gamma (N)}\int_{m^{2}}^{s}du(s-u)^{N-1}f(u),\, \,\,N>0.
\end{equation}%
For $N\leq 0$ we have also used:
\begin{eqnarray}
&&\widetilde{F}_{N}(s)=\left( \frac{d}{ds}\right) ^{-N}\left[
f(s)\int_{1}^{\infty }\frac{dt}{t}\exp \left( -\frac{\Lambda ^{2}t}{s-m^{2}}%
\right) \right], \notag \\
&&\widetilde{F}_{N}(s_{0})=\left( \frac{d}{ds_{0}}\right) ^{-N}\left[
f(s_{0})\ln \left( \frac{s_{0}-m^{2}}{\Lambda ^{2}}\right) -\gamma
_{E}\right]. \notag \\
&&{}
\end{eqnarray}%
The expression provided above are valid only if $f(m^2)=0$. In other cases,
one has to use the prescription $f(s)=[f(s)-f(m^2)]+f(m^2)$, where the first term in
the brackets is equal to zero, when $s=m^2$.

\end{document}